# A quantitative approach to choose among multiple mutually exclusive decisions: comparative expected utility theory

Zhu, Pengyu

[Cluster on Industrial Asset Management, University of Stavanger, N-4036, Stavanger, Norway]

## Abstract

Mutually exclusive decisions have been studied for decades. Many well-known decision theories have been defined to help people either to make rational decisions or to interpret people's behaviors, such as expected utility theory, regret theory, prospect theory, and so on. The paper argues that none of these decision theories are designed to provide practical, normative and quantitative approaches for multiple mutually exclusive decisions. Different decision-makers should naturally make different choices for the same decision question, as they have different understandings and feelings on the same possible outcomes. The author tries to capture the different understandings and feelings from different decision-makers, and model them into a quantitative decision evaluation process, which everyone could benefit from. The basic elements in classic expected utility theory are kept in the new decision theory, but the influences from mutually exclusive decisions will also be modeled into the evaluation process. This may sound like regret theory, but the new approach is designed to fit multiple mutually exclusive decision scenarios, and it does not require a definition of probability weighting function. The new theory is designed to be simple and straightforward to use, and the results are expected to be rational for each decision-maker.

Keywords: mutually exclusive decisions; decision theory; comparative cost of chance; comparative expected utility theory; mathematical analysis; game theory; probability weighting;

## 1. Introduction

Traditionally, decision theories could be divided into descriptive, normative and prescriptive methods, and different methods featured different assessment processes or tools (Bell & Raiffa, 1988; Bell, Raiffa, & Tversky, 1988; Luce & Von Winterfeldt, 1994). Descriptive decision theories are concerned to explain 'how people make decisions', rationally, bounded rationally, and irrationally (Keeney, 1992, pp. 57,58). Typical descriptive decision theories include prospect theory, naturalistic theory, image theory, and so on. These theories normally featured with conclusions from experiments on human behaviors under risks and uncertainties (Bell, 1985; Tversky & Kahneman, 1981). Normative decision theories focus on 'rational procedures for decision making', such as expected utility theory, and regret theory (Bermúdez, 2009; Keeney, 1992, pp. 57,58). The connection between normative assessment and rationality was discussed by Bermúdez (2009). The purpose of normative modelling of decision problem is to benefit those who are more rational during decision making procedure (Bermúdez, 2009; Tversky & Kahneman, 1986). Prescriptive decision theories are focused on applications of descriptive and normative approaches to help people make informed decision in practice, such as analytical hierarchy process (Keeney, 1992, pp. 57,58).

Decision among multiple mutually exclusive options means that one is only allowed to choose one from all the options. For mutually exclusive decisions, many well-known decision theories have been defined to either help people make rational decisions, or to interpret people's behaviors, such as expected utility theory (EUT), regret theory, prospect theory and so on. Utility is defined as some measurable 'immediate sensation' with no need of further analysis (Von Neumann & Morgenstern, 1944, p. 16). For classic expected utility theory, several axioms have been defined, including transitivity of preference, dominance/maximization, combinability and continuity, independence/cancellation, and invariance (Friedman & Savage, 1948; Tversky & Kahneman, 1986; Von Neumann & Morgenstern, 1944). However, these fundamental principle of EUT have been challenged by many researchers. The famous Allais paradox experiment has challenged the dominance axiom of EUT and the Savage independence axiom through experiments (Allais, 1953). Ellsberg (1961) also observed possible violation of Savage independence axiom of EUT. In his setting, two different decision pairs shall become identical if independence axiom was applied, which is contradictive with the experiment results. Kahneman and Tversky (Tversky & Kahneman, 1981; Tversky & Kahneman, 1986, 1987) have made several observations of direct violations of transitivity of preference axiom, dominance axiom, and invariance axiom of EUT. Many observations have challenged the validity of EUT of being a normative decision-making method.

Some theorists tried to explain these phenomenon by embedding psychological aspects into decisions. There were long discussions and explorations on the possibility to merge normative method with descriptive method (Bell, 1985; Bell et al., 1988; Kahneman & Tversky, 1979; Tversky & Kahneman, 1987). The process was quite fruitful and some theories like prospect theory and regret theory were suggested (Bell, 1982, 1985; Fishburn, 1982; Kahneman & Tversky, 1979; Loomes & Sugden, 1982). Both theories have been widely discussed and implemented in many fields (Bleichrodt, Pinto, & Wakker, 2001; Camerer, 2004; Filiz-Ozbay & Ozbay, 2007; Quiggin, 1994; Taylor, 1997). Prospect theory uses EUT as analysis basis and includes psychological aspects in interpreting people's behaviors. Prospect theory could provide good explanations to several contradictive observations based on expected utility theory. However, prospect theory is more like descriptive theory other than normative theory. The main focus of the theory is on subjective probability weighting, and the evaluation process does not directly include other mutually exclusive decisions. Regret theory established a normative approach to include exclusive decision options into calculation, but study on this theory is limited on pairwise choices, and the results were not transitive (Bell, 1982; Loomes & Sugden, 1982). Non-transitivity means that the theory should never fit decisions that have more than two options, as there might be no answer to be suggested. In addition, all these theories require some kind of individual definition of psychological factors, such as regret-rejoice function and probability weighting function. This is very inconvenient for practical application. It is observed that there is not such a theory that provides practical and normative procedure for multiple mutually exclusive decisions, and at meantime reflects the possible psychological aspects from different decision-makers.

In this paper, the author will explain which parameter could be added and calculated to make such comparisons happen. Some new concepts will be introduced, and a detailed process of applying the alternative decision theory will be explained.

## 2. What is opportunity cost for mutually exclusive decisions

Cost does not necessarily mean economic cost, while economic cost is simple to explain and straightforward to understand. For mutually exclusive decisions, different choices are not independent with each other, and one choice's value could be an opportunity cost for the other exclusive choice. In economics, opportunity cost is included during evaluation of decision choices. It represents the loss of potential opportunities from the best mutually exclusive choice that one does not choose (Jaffe & Randolph Westerfield, 2004; McWatters & Zimmerman, 2015; Ross, 2007). One of Kahneman and Tversky's experiments could be used to explain the opportunity cost from potential exclusive alternatives, as described in Table 1. The comparison is done with expected utility theory.

*Table 1 Choice of gamble (Kahneman & Tversky, 1979)*

| Choice A | Probability | 80% | 20% | Choice B | Probability | 100% |
|---|---|---|---|---|---|---|
| | Outcome | 4000 | 0 | | Outcome | 3000 |

In the decision scenario in Table 1, the expected outcome of choice A is $Ex_A$= 4000*80%=3200 and the expected outcome of choice B is $Ex_B$= 3000*100%=3000. If opportunity cost is considered, $Ex_B$ is the opportunity cost for the decision-maker if A is chosen, and vice versa. The conventional way to include opportunity cost during decision-making is done by directly deducting the expected value from the second best alternative decision option that could have been chosen. In this example, if time aspect is not taken into account, the decision between the two choices with consideration of opportunity cost thus become choices between 3200-3000 (choosing A) and 3000-3200 (choosing B). This gives some insights on the potential values that one might miss by choosing another alternative option.

Opportunity cost is identified as the second best exclusive decision apart from the one that is to be chosen, and the evaluation of different decision options is just to deduct the opportunity cost from the value of the chosen decision. This definition determines that the decision with higher value will be balanced with a lower opportunity cost, and the decision with lower value will be deducted with a relatively higher opportunity cost. It could be calculated that inclusion of opportunity cost will never change the order of decisions, when traditional EUT is used.

### 2.1. Subjectivity of defining influence from mutually exclusive decisions

The fact that opportunity cost would never change the priority of decisions does not sound logical, even though this practice has been around for a long time. Mutually exclusive decisions are not independent, and should have influences on each other as well as final preferences. Opportunity cost is often used in economic domain. In decision-making domain, the influence from other mutually exclusive decisions has been modelled as regret and rejoice through Loomes and Sugden's regret theory.

It has long been observed that different decision makers treat the same decision questions in different ways. Traditional explanations on these observations may include risk profile definitions, probability weighting difference, different interpretations of the same problem, and so on (Bleichrodt et al., 2001; Tversky & Kahneman, 1981). Similarly, the definition of regret should vary from one decision-maker to another. How to model this subjectivity and influence from exclusive decisions has been challenging. Regret theory suggested a regret-rejoice function, but the theory could only apply on pairwise comparisons and transitivity is not applicable. A solution for multiple mutually exclusive decisions needs to be further explored.

Even though Loomes and Sugden's regret theory does not seem to be applicable in multiple mutually exclusive decisions conditions, the inclusion of regret/rejoice in evaluating mutually exclusive decisions is still inspiring. The theory tried to model those impacts into a normative evaluation process, which is also appreciated by the author. No matter how the influence from the mutually exclusive decisions is named, the individual subjectivities need to be pointed out and modeled as they genuinely should be.

### 2.2. Comparative cost of chance

If a closer look is taken into the decision scenario in Table 1, which condition would make a gambler feel truly regret if he/she chooses one instead of the other? If A is chosen, there is 20% chance that he/she will get nothing. It is reasonable to say that this 20% probability is the part that might make most people feel sorry for, as the other outcomes seem to be good enough (subjective). This 20% also means the potential loss is $Ex_B$=3000, as if B has been chosen. The gambler will not feel sorry about the potential gain of 3000 if he/she could get the luck to win 4000 (80% chance). Similarly, the gambler who chooses B will get a 100% guarantee to win 3000 without any chance of unexpected outcome, if he/she is happy about 3000 instead of possible 4000 with 80% probability. The regret or opportunity cost for choosing A instead of B is thus calculated as 20%*$Ex_B$=600, and the regret or opportunity cost for choosing B instead of A is 0%*$Ex_A$= 0.

The above analysis provides another perspective of defining the opportunity cost or regret during the evaluation of mutually exclusive decisions. All possible outcomes of mutually exclusive decisions and their associated probabilities need to be considered in this case. The author believes that definitions of opportunity cost need to be done on the level that each possible outcome should be evaluated and marked as regretful or not. In addition to this, the author believed that there should be no need to assign probability weighting functions or any regret factors during the evaluation of different decision options, as the probability of occurrence associated with each possible outcome could be used for evaluation.

This is not the traditional way to include opportunity cost or regret into calculation, as it differentiates the probability of having that opportunity cost or regret. In order to reduce possible confusion, the author describes the influences from mutually exclusive decisions as comparative cost of chance, or CCC in short.

'Comparative' indicates that application of such cost is about comparison between mutually exclusive decision options. In the following section, the author will propose a new decision framework that is based on the expected utility theory and takes into account of comparative cost of chance.

## 3. An alternative approach to choose among multiple mutually exclusive decisions

Before further discussion, it should be noted that some decisions can be made with simple comparisons of expected value and variance for risk-neutral decision makers. For example, if choice A has higher expected value and lower variance, A should be a reasonable choice to be picked and no further analysis should be needed. In addition, decisions that allow combination of choices are not in the discussion scope for this article, and methods to aid decision-making under that condition could be found in other researches (Lang & Stulz, 1993; Lintner, 1965; Rumelt, 1982). The discussions of alternative decision theory are focused on mutually exclusive decisions.

### 3.1. Identification and calculation of Comparative Cost of Chance

As discussed in Section 2.2, the definition of opportunity cost in this paper is different from traditional so-called opportunity cost, and is named as comparative cost of chance (CCC). In brief, for any mutually exclusive decision alternative, CCC refers to the possible decision outcome that would make certain decision-makers feel unexpected or sorry about. In a gaining condition (money, or other utility), the unexpected part could be lower gain or loss compared to other alternatives. In a losing or spending condition, the unexpected part could be more loss compared to other options. Those outcomes and their associated probabilities that are considered as regretful are marked as CCC elements. The identification of CCC elements is suggested to follow dominance axiom of expected utility theory. However, the process is subjective, and different decision-makers may define CCC in drastically different ways. Decisions are influenced by definitions of CCC elements.

Since the identification of CCC elements is subjective, the results should also reflect the decision-makers' risk profiles and preferences. For example, some small loss might be accepted by people with risk-seeking profiles, and not considered as a CCC element. During the identification of CCC process, some values are obvious to be CCC element for many decision-makers, and some values maybe hard to tell. Decision makers get to decide which elements would or would not satisfy their expectations. For instance, in Table 1, if B is chosen, there is 100% to win 3000. For some people, 3000 is a good prize and leaves no reason to feel sorry about. The possible outcome should not be identified as a CCC element for these people. This also means that there is 0% to feel sorry about anything. CCC is calculated as $0\%*Ex_A=0$. However, for another group of people, 3000 might not be high enough and the possibility of winning a higher prize with a decent probability from another decision alternative make them feel regretful for choosing B. This may sound irrational, but different decision-makers ought to choose whatever suit their purposes. This outcome thus become a CCC element for the latter group of people. Since the probability of this outcome is 100%, the CCC for this group of people is calculated as $100\%*Ex_A=3200$.

### 3.2. Inclusion of Comparative Cost of Chance into calculation

The example explains how CCC could be identified in a gaining condition (expected values are positive). For losing conditions, the identification and calculation of CCC are similar. The principle of identifying CCC elements is still to select the element that makes certain decision maker feels sorry about, that is to say, lose less money, for instance. What is changed is the inclusion of CCC into calculation in losing conditions. In a gaining condition, CCC is deducted from option's expected value to reflect the negative influence from other decision options. In losing conditions, the possible less loss from other exclusive decisions by choosing certain decision option marks CCC elements. Since it is a losing condition, both expected values and CCC are negative. If one feels regretful for choosing one option, the CCC should not be deducted from current option's expected value (like what is did in gaining conditions), because this will make the negative loss become smaller as a matter of mathematical fact, which is not logical. The possibility to lose less (by choosing another exclusive decision option) make the current option's loss more painful. The actual CCC should be added to decision option's expected value so that a greater loss feeling from decision-makers can be modelled in a meaning way.

### 3.3. The theory of comparative expected utility

With the introduction of CCC, a new approach for decision-making could be suggested. The alternative decision theory uses expected utility theory (EUT) as a comparison bases. The theory includes the impacts of other mutually exclusive decisions into evaluation as CCC elements. Since this decision theory is based on EUT and CCC, it gets a name as **Comparative Expected Utility Theory**. The definition of the theory is defined as:

*The comparative expected utility theory (CEUT) is introduced as a normative method to help decision makers with all kinds of risk profiles assess multiple mutually exclusive decision alternatives. CEUT method uses the results from classic expected utility theory as bases, and takes into consideration of comparative cost of chance (CCC), which is a function of regretful expected outcome from another decision alternative and the probability of that regretful outcome to happen. CEUT method is both quantitative and subjective.*

The application of CEUT method for multiple mutually exclusive decisions could be implemented in five steps:

Step 1. Modeling of decision options;
Step 2. Calculation of expected utility;
Step 3. Identification of elements that is considered to be comparative cost of chance (CCC);
Step 4. Calculation of $CCC_A = \text{sum}(p(CCC_A)) * Ex_B$,
where $\text{sum}(p(CCC_A))$ is the summary of probability of decision option A's all CCC elements, and $Ex_B$ is the best option among the unchosen decision alternatives;
Step 5. Calculation of comparative expected utility (CEU)= expected utility-|CCC|,
where |CCC| is the absolute value of CCC.;

Based on the discussions in Section 3.2, the inclusion of CCC values into calculation vary from gaining conditions to losing conditions. In order to make a generalized process that suit both gaining and losing conditions, the absolute value of CCC is used for the calculation of comparative expected utility value. This explains why an absolute value of CCC is used in Step 5. The simplified function for calculation comparative expected utility of certain decision option is: **CEU=Ex-|CCC|.**

### 3.4. Application of CEUT method

If classic expected utility is compared between the two choices in the example in Table 1, A would be recommended over B. However, in the original experiment, 80% of participants chose B over A (Kahneman & Tversky, 1979). It is very likely that people would still choose B over A with the knowledge that A has a higher expected value. It is interesting to see what CEUT method suggests in the same decision question. The five-step procedure is followed:

Step 1. Modelling of decision question, as shown in Table 1.
Step 2. Calculation of expected utility

The utility is interpreted as monetary value in this example. The expected value for two choices are: $Ex_A$=80%*4000+20%*0=3200, and $Ex_B$=100%*3000=3000;

Step 3. Identification of CCC elements

The identification of CCC elements is subjective based on different individual/organizations' preferences and their risk profiles. In general, all the possible outcomes could be evaluated against the best alternative decision in terms of expected value. The evaluation principle may vary from one decision-maker to another. In this example, the evaluation principle of the group of people in the original experiment is defined based on their actual behaviors.

For choice A, '80% to get 4000' is definitely appreciated compared to the expected outcome from choice B, as the possible outcome is much higher than the other choice's expected value; '20% chance to get 0' should thus be identified as a CCC element. For choice B, the only choice is 100% to get 3000. Most people prefer choice B in the original experiment, and the majority of the group of people thus appeared to be risk-averse. It is reasonable to assume that this group of people would not consider '100% to win 3000' to be regretful. In other word, there is 0% probability to feel regret about the possible outcome of choice A for this group of people. All CCC elements are marked with 'x', as shown in Table 2. In this example, only one possible outcome in choice A has been identified as CCC element.

*Table 2 Identification of CCC elements, adjusted from experiment from Kahneman and Tversky (1979)*

| Choice A | Probability | 80% | 20% | Choice B | Probability | 100% |
|---|---|---|---|---|---|---|
| | Outcome | 4000 | 0 | | Outcome | 3000 |
| | CCC elements | - | x | | CCC elements | - |

Step 4. Calculation of CCC

For this group of decision-makers, the comparative cost of chance by choosing A is $CCC_A = \text{sum}(p(CCC_A)) * Ex_B$, where $\text{sum}(p(CCC_A))$=20%, as marked in Table 2 . $CCC_A = \text{sum}(p(CCC_A)) * Ex_B$=20%*3000=600.

No CCC element is identified in choice B, so $CCC_B=0$.

In this case, only 2 decision options are given, but similar process could be done on multiple mutually exclusive decisions. In those cases, each possible outcome will need to be evaluated against the expected value of the best alternative decision among the rest that are not chosen. CCC for one decision option will be calculated with the summary of probability of all CCC elements in this option and the expected value of the best alternative option.

Step 5. Calculation of CEU

According to the definition of comparative expected theory, for choice A, $CEU_A = Ex_A - |CCC_A| = 3200 - 600 = 2600$.

For choice B, $CEU_B = Ex_B - |CCC_B| = 3000 - 0 = 3000$.

The axioms of CEUT has not been explained yet. However, since CEUT is a quantitative approach, the results should be numerically comparable, and dominance axiom should apply in CEUT. Based on this assumption, Choice B has a higher CEUT value than choice A. B is suggested and should be chosen this group of decision makers.

The CEUT result actually aligns with the observations from the original experiment. This observation is promising for CEUT method. However, the axioms of the theory need to be defined and validated.

## 4. Evaluation of possible axioms of CEUT

CEUT is designed to be used for multiple mutually exclusive decision options. CEUT uses the inputs of EUT as comparison bases and introduces comparative cost of chance into evaluations. The axioms of EUT are hereby evaluated and validated for fitness of CEUT method. In the above example, dominance axiom was assumed to fit for CEUT approach when comparing different alternatives. This need to be further studied and validated. In addition, transitivity of preference, combinability and continuity, independence/cancellation, and invariance axioms from EUT also need to be studied (Friedman & Savage, 1948; Tversky & Kahneman, 1986; Von Neumann & Morgenstern, 1944).

### 4.1. Dominance or maximization axiom

CEUT provides a quantitative approach to assess mutually exclusive decisions, and the results should naturally be numerically comparable. If the first choice's CEU value is greater than the second one's, the first one should be recommended compared to the second one if the dominance axiom stands.

The CEU results on the example in Table 1, as calculated in Section 3.3, have shown good alignment of CEUT results and people's actual decisions, based on the dominance axiom of the theory. In another experiment from Kahneman and Tversky (1979), a loss scenario showed that people tended to behave in a risk-seeking manner, even though the potential loss from certain choice could be higher than other choices. The experiment and the application of CEUT on it are shown in Table 3.

*Table 3 Choice of gamble, adapted from Kahneman and Tversky (1979)*

| Choice A | Probability | 100% | Choice B | Probability | 75% | 25% |
|---|---|---|---|---|---|---|
| | Outcome | -750 | | Outcome | -1000 | 0 |
| | $Ex_A$ | -750 | | $Ex_B$ | -750 | |
| | CCC element | x | | CCC element | x | - |
| | $|CCC_A|$ | 750 | | $|CCC_B|$ | 562.5 | |
| | $CEU_A$ | -1500 | | $CEU_B$ | -1312.5 | |

The identification of CCC elements in this case is based on the logic from the original experiment from Kahneman and Tversky (1979). For that group of people, the certainty (100% probability) to lose 750 was disliked by the majority of group, and a slight chance to lose nothing was appreciated. CCC elements of the two decision choices are identified accordingly when CEUT approach is applied. The results from Table 3 shows that choice B has a higher CEU value than choice A. The original experiment also gave the same results. This is not a coincidence, as the implementation of CEUT on this example is carried out based on the preferences from the same group of people. Unlike other psychological and qualitative analyses, CEUT provides decision-makers a normative process to evaluate different decisions and a quantitative approach to compare the results. Dominance axiom serves as a basic axiom in CEUT.

### 4.2. Transitivity of preference

For transitivity of perference, when u is preferred over v and v is preferred over w, u should be preferred over w (Von Neumann & Morgenstern, 1944). The preference relationship could be expressed by using the conventional notation >, i.e. when u>v and v>w, u>w. In EUT, the preference is transitive. The transitivity axiom is one of the reasons that the author suggests the CEUT method instead of using regret theory, where transitivity does not apply. In CEUT, the quantity nature does not change with the introduction of CCC elements. Since dominance axiom applies in the theory, transitivity axiom should also apply in CEUT.

For example, a gambler is facing three choices of games, and he/she is only allowed to choose one of them. The choices are listed in Table 4.

*Table 4 Choice of gamble, choice A and choice B are adapted from Kahneman and Tversky (1979)*

| Choice A | Probability | 80% | 20% | Choice B | Probability | 100% | | Choice C | Probability | 80% | 20% |
|---|---|---|---|---|---|---|---|---|---|---|---|
| | Outcome | 4000 | 0 | | Outcome | 3000 | | | Outcome | 5000 | 0 |
| | Ex | 3200 | | | Ex | 3000 | | | Ex | 4000 | |

Before CEUT is applied, the preferences on certain gains or losses of decision makers need to be clarified. This has a direct influence on the identification of CCC elements, which might lead to different decision recommendation. In this example, it is assumed that the gambler will not feel regret if he/she chooses choice B of '100% to win 3000' even though there is a chance to get 4000 or 5000 in other games. This outcome will not be identified as a CCC element for these group of decision makers. The procedures of applying CEUT are simplified, and only results are used.

If pairwise comparison manner is used on options shown in Table 4:

- Between A and B: $CEU_A = Ex_A - |CCC_A| = 3200 - 20\% \ast 3000 = 2600$; $CEU_B = Ex_B - |CCC_B| = 3000$. B > A.
- Between B and C: $CEU_B = 3000$; $CEU_C = 4000 - 20\% \ast 3000 = 3400$. C > B.
- Between A and C: $CEU_A = 3200 - 20\% \ast 4000 = 2400$; $CEU_C = 4000 - 20\% \ast 3200 = 3360$. C > A;

If transitivity is applicable in CEUT, B>A and C>B should imply C>A. It is seen that CEUT results also imply C>A. Transitivity axiom seems to be valid with CEUT approach.

If the three choices are to be compared at the same time, the results should stay the same.. When more than two decisions are evaluated, all possible outcomes will need to be evaluated against the expected value from the best alternative decision. The best alternative option could be different for different decision choices. If A or B is to be chosen, the best alternative choice is C due to its high expected value. If C is chosen, the best alternative choice will be A due to its higher expected value. The simplified calculation process with CEUT is shown in Table 5

*Table 5 Choice of gamble, choice A and choice B are adapted from Kahneman and Tversky (1979)*

| Choice A | Probability | 80% | 20% | Choice B | Probability | 100% | | Choice C | Probability | 80% | 20% |
|---|---|---|---|---|---|---|---|---|---|---|---|
| | Outcome | 4000 | 0 | | Outcome | 3000 | | | Outcome | 5000 | 0 |
| | $Ex_A$ | 3200 | | | $Ex_B$ | 3000 | | | $Ex_C$ | 4000 | |
| | CCC elements | - | x | | CCC elements | - | | | CCC elements | - | x |
| | $|CCC_A|$ | - | 20%*4000 | | $|CCC_B|$ | - | | | $|CCC_C|$ | - | 20%*3200 |
| | $CEU_A$ | 2400 | | | $CEU_B$ | 3000 | | | $CEU_C$ | 3360 | |

From the results, $CEU_C > CEU_B > CEU_A$. In other words, C> B> A. This results is consistent with the results from the pairwise comparison approach with CEUT. Transitivity of preference is considered as an important part of rationality (Schauenberg, 1981; Von Neumann & Morgenstern, 1944), and the transitivity axiom of CEUT makes it possible to rationalize decision-making process as well as to predict decisions in many occasions.

### 4.3. Combining and continuity axiom

CEU theory is designed for mutually exclusive decision choices. Combining two decision choices is out of the discussion scope in this context. Combining and continuity axiom does not apply in CEU theory.

### 4.4. Independence/cancellation axiom

According to EUT, independence/cancellation axiom exists provided that the different decision alternatives are independent from one another (Von Neumann & Morgenstern, 1944). Violation of the independence axiom of EUT has been observed (Allais, 1953; Ellsberg, 1961; Zhou, 2004). According to CEUT, different decision alternatives are not independent as they are mutually exclusive. In addition, the probability associated with each possible decision outcome might be used in the calculation of CEU results. Independence/cancellation axiom is contradictive with CEUT procedures in nature and should not apply in CEUT.

The well-known Allais paradox could be used to explain the incompatibility. However, the difference between two decision options in the original setting is relatively big. 500 million in the original setting is here changed to 115 million (could be any number). The new gamble game (two decisions: s1 vs r1 and s2 vs r2) is described in Table 6.

*Table 6 Choice of gamble (unit: million), modified from Allais (1953)*

| Choice s1 | Probability | 100% | Choice r1 | Probability | 10% | 89% | 1% |
|---|---|---|---|---|---|---|---|
| | Outcome | 100 | | Outcome | 115 | 100 | 0 |

| Choice s2 | Probability | 11% | 89% | Choice r2 | Probability | 10% | 90% |
|---|---|---|---|---|---|---|---|
| | Outcome | 100 | 0 | | Outcome | 115 | 0 |

A simplified procedure of applying CEUT on the example and the results are shown in Table 7.

*Table 7 Comparison of choices using CEU theory*

| Choice s1 | Probability | 100% | | Choice r1 | Probability | 10% | 89% | 1% |
|---|---|---|---|---|---|---|---|---|
| | Outcome | 100 | | | Outcome | 115 | 100 | 0 |
| | $Ex_{S1}$ | 100 | | | $Ex_{r1}$ | | 100.5 | |
| | CCC elements | - | | | CCC elements | - | - | x |
| | $|CCC_{S1}|$ | - | | | $|CCC_{r1}|$ | - | - | 1 |
| | $CEU_{S1}$ | 100 | | | $CEU_{r1}$ | | 99.5 | |

| Choice s2 | Probability | 11% | 89% | Choice r2 | Probability | 10% | 90% |
|---|---|---|---|---|---|---|---|
| | Outcome | 100 | 0 | | Outcome | 115 | 0 |
| | $Ex_{S2}$ | 11 | | | $Ex_{r2}$ | 11.5 | |
| | CCC elements | - | x | | CCC elements | - | x |
| | $|CCC_{S2}|$ | 10.235 | | | $|CCC_{r2}|$ | - | 9.9 |
| | $CEU_{S2}$ | 0.766 | | | $CEU_{r2}$ | 1.6 | |

The results show that s1 has higher CEUT value than r1, and r2 has higher CEUT value than s2. In brief, s1>r1, and s2<r2.

If cancellation axiom is applicable in CEUT, 89% to win 100 is canceled in both s1 and ri, and 89% to win 0 is canceled in both s2 and r2. The two choices will become exactly identical, as shown in Table 8. This means that if s1>r1, s2 should also be preferred over r2. This is not consistent with the results from CEUT.

*Table 8 Choices if using cancellation axiom*

| Choice s1 | Probability | 11% | Choice r1 | Probability | 10% | 1% |
|---|---|---|---|---|---|---|
| | Winning | 100 | | Winning | 115 | 0 |

| Choice s2 | Probability | 11% | Choice r2 | Probability | 10% | 1% |
|---|---|---|---|---|---|---|
| | Winning | 100 | | Winning | 115 | 0 |

The reason for this is that CEUT needs to take probability of CCC elements into calculation. Application of independence axiom will change the function of CEUT, and the result will be changed accordingly. Independence/cancellation axiom does not apply in CEU theory.

### 4.5. Invariance axiom

Kahneman and Tversky (1984) have challenged the axiom of invariance in EUT. It was observed that different expressions of the same problem influenced decisions from the same group of people (Kahneman & Tversky, 1984). The original experiments were modelled in Table 9. Positive outcome in the table means life savings and negative outcome refers to life losses.

Table 9 The patient example, adapted from Kahneman and Tversky (1984)

| Choice A | Probability | 100% | | Choice B | Probability | 1/3 | 2/3 |
|---|---|---|---|---|---|---|---|
| | Outcome | 200 | | | Outcome | 600 | 0 |
| Choice C | Probability | 100% | | Choice D | Probability | 1/3 | 2/3 |
| | Outcome | -400 | | | Outcome | 0 | -600 |

The decision between A and B and the decision between C and D are two different ways to express the same decision question. The first one focuses on life savings from certain decision, and the latter one is concerned with life losses. The original experiment results showed that the same group of people chose A over B, and chose D over C. The experiment results revealed that the same group of people might be both risk-averse (in gaining/saving condition) and risk-seeking (in losing condition) when the same decision question is framed in different ways (Bell, 1985; Kahneman & Tversky, 1984). The invariance axiom of EUT was challenged.

The observation on people's different attitudes on gaining and losing has become one of the fundamental elements for decision theories like regret theory, and prospect theory (Bell et al., 1988; Tversky & Kahneman, 1992; Zhou, 2004). However, CEUT, as a quantitative and normative method, is designed to give recommendations based on quantitative results. The way that the decision problem is addressed should not become a reason that the recommendations would ever be changed. The invariance axiom should also apply in CEUT, or else transitivity axiom and dominance axiom would not stand either. The invariance axiom will be tested with the famous experiment that is shown in Table 9.

In order to use CEUT on this example, it is necessary to assume that this group of decision makers should keep the consistency of subjective preference. Based on the results from the original experiment, 72% of the group of people chose A over B (Kahneman & Tversky, 1984). For this group of people, '100% to save 200 lives' should not be identified as a CCC element compared to the option B. In the experiment setting, saving 200 lives also means the rest of the patients (400 out of 600) would not be saved. Since the outcome in choice A is not identified as a CCC element, 'The 100% to lose 400 lives' in choice C should not be identified as a CCC element either, if the condition is fully understood by the group of people. Based on these assumptions, CEUT is applied on the example, and the results are shown in Table 10.

Table 10 Implementation of CEUT, original data from Kahneman and Tversky (1984)

| Choice A | Probability | 100% | | Choice B | Probability | 1/3 | 2/3 |
|---|---|---|---|---|---|---|---|
| | Outcome | 200 | | | Outcome | 600 | 0 |
| | $Ex_A$ | 200 | | | $Ex_B$ | 200 | |
| | CCC elements | - | | | CCC elements | - | x |
| | $|CCC_A|$ | - | | | $|CCC_B|$ | - | 133 |
| | $CEU_A$ | 200 | | | $CEU_B$ | 67 | |
| Choice C | Probability | 100% | | Choice D | Probability | 1/3 | 2/3 |
| | Outcome | -400 | | | Outcome | 0 | -600 |
| | $Ex_C$ | -400 | | | $Ex_D$ | -400 | |
| | CCC elements | - | | | CCC elements | - | x |
| | $|CCC_C|$ | - | | | $|CCC_D|$ | - | 267 |
| | $CEU_C$ | -400 | | | $CEU_D$ | -667 | |

From the CEUT results, A>B, and C>D. Decision recommendations are not influenced by the different ways of framing the same decision question, when CEUT is applied. The above CEUT application is based on that '100% to save 200 lives or 100% to lose 400 lives' is not considered as a CCC element, as what is reflected from the original experiment. For another group of people, the option might be considered as a CCC element. However, the consistency should still be kept. The calculation is skipped here, and the recommendation of choice will become B>A, and D>C. The results from CEUT keep a good consistency regardless of the way that decision question is framed. Invariance axiom applies in CEU theory.

## 5. Is CEUT method applicable on outcomes with small probabilities

It could be seen that applications of CEUT are very convenient and reliable through the previous discussions. There is another situation that needs to be further addressed and tested, which is outcome with small probabilities or rare events. Overweighting of small probabilities from decision-makers has been constantly observed (Gonzalez

& Wu, 1999; Kahneman & Tversky, 1979). If EUT is not to be trusted, people often find themselves clueless on how these outcomes with small probabilities should be evaluated. It is also noticeable that rare events also come with either very small or very big values. The combination of very small or large values with rare probabilities makes the evaluation even more difficult. This might be good news for insurance companies or betting companies, whose business model is mainly based on such events.

The Problem 14 from Kahneman and Tversky (1979) is used to test CEUT method on such events. The original experiment is described in Table 11.

*Table 11 Choice of gamble (Kahneman & Tversky, 1979)*

| Choice A | Probability | 0.1% | 99.9% | Choice B | Probability | 100% |
|---|---|---|---|---|---|---|
|  | Outcome | 5000 | 0 |  | Outcome | 5 |

In the original experiment, 72% of participants chose A over B (Kahneman & Tversky, 1979). The expected value for choice A and choice B are both 5. EUT principle could not be used to tell the preference between the two choices. However, the lower variance of choice B compared to choice A makes B a better choice with traditional mindset. The observations from the experiment did not support this logic, though. Kahneman and Tversky (1979) explained that the results are derived from overweighting of occurrence of rare events. It is interesting to see what CEUT method suggests in the same decision scenario.

The identification of CCC elements is subjective based on different individual/organizations' preferences and their risk profiles. For choice B, the only choice is 100% to get 5. It is possible that some decision-makers may define it as a CCC element and some may not. In the original experiment from Kahneman and Tversky, most of the experiment attendances chose '0.1% to win 5000' over '100% to win 5'(Kahneman & Tversky, 1979). It was explained that most people might overweight small probabilities and underweight large probabilities (Kahneman & Tversky, 1979). Loomes and Sugden (1982) interpreted this observation just like some people's willingness to buy a lottery ticket for big prize with small probability. For this specific group of people from the original experiment, choosing '100% to win 5' let them lose the chance to win a big prize (mutually exclusive decisions) and this choice should be marked as a CCC element.

The simplified process and calculation with CEUT are as shown in Table 12.

*Table 12 Identification of CCC elements, adjusted from experiment from Kahneman and Tversky (1979)*

| Choice A | Probability | 0.1% | 99.9% | Choice B | Probability | 100% |
|---|---|---|---|---|---|---|
|  | Outcome | 5000 | 0 |  | Outcome | 5 |
|  | $Ex_A$ | 5 |  |  | $Ex_B$ | 5 |
|  | CCC elements |  | x |  | CCC elements | x |
|  | $|CCC_A|$ | 4.995 |  |  | $|CCC_A|$ | 5 |
|  | $CEU_A|$ | 0.005 |  |  | $CEU_B|$ | 0 |

From the calculation, $CEU_A > CEU_B$, and A>B. This result is consistent with the original experiment observation. This also explains why many people could easily buy a lottery ticket without worrying about the almost assured loss of the money that is paid on the lottery.

In the original experiment, a symmetric experiment (Problem 14') was designed with losing condition (Kahneman & Tversky, 1979). The decision question is shown in Table 13. It is interesting to compare the experiment results between these two decisions. The setting of the two experiments were totally symmetrical, but the results were the opposite. The same group of people weighted gaining and losses in different scales and reflected two risk profiles in gaining and losing scenarios (Kahneman & Tversky, 1979).

*Table 13 Choice of gamble (Kahneman & Tversky, 1979)*

| Choice C | Probability | 0.1% | 99.9% | Choice D | Probability | 100% |
|---|---|---|---|---|---|---|
|  | Outcome | -5000 | 0 |  | Outcome | -5 |

When CEUT is applied with the experiment. If choice C is chosen, the 0.1% of getting -5000 should be identified as a CCC element. 99.9% to lose nothing should not be identified as a CCC element as the alternative is to lose 5 with 100% probability.

For choice D, since the same group of people in earlier experiment (Table 12) treated small loss of monetary value as not regretful. The certainty to lose 5 should not make a big difference for them either. That means they would 'gladly' accept this loss of 5 just like that they did not want to win just 5 with 100% probability. This element of '100% to lose 5' is thus not identified as a CCC element. All CCC elements are marked with 'x' and simplified CEUT calculation is shown in Table 14.

*Table 14 Calculation of CEU, adjusted from Kahneman and Tversky (1979)*

| Choice C | Probability | 0.1% | 99.9% | Choice D | Probability | 100% |
|---|---|---|---|---|---|---|
| | Outcome | -5000 | 0 | | Outcome | -5 |
| | Ex | -5 | | | Ex | -5 |
| | CCC elements | x | - | | CCC elements | - |
| | $|CCC_C|$ | 0.005 | 0 | | $|CCC_D|$ | - |
| | $CEU_C$ | -5.005 | | | $CEU_D$ | -5 |

From the calculation, $CEU_C < CEU_D$, and C<D. This might be understood by the fact that many people choose to pay insurance company for rare hazardous events.

Unlike the descriptive explanations from other researchers, CEUT could provide a quantitative and normative approach to implement such comparisons. The results from CEUT also showed great consistency with experiment observations.

The application of CEUT on the above two examples are based on the assumptions that certain group of people do not appreciate small gains or losses. For another group of people, they might consider gaining 5 for certain is not regretful compared to a lottery ticket, and they might not think losing 5 for certain is ok if they would have known there is a 99.9% probability of losing nothing. If CEUT is applied on these two examples for this group of decision-makers, the results would suggest B over A (B>A), and recommend C over D (C>D). The calculation is skipped here. If this decision is explained in term of lottery ticket and flight insurance as suggested by Loomes and Sugden (1982), the preference of B and C based on CEUT results reflects the group of people who do not buy lottery and the group of people who do not buy flight insurance. In other words, if certain group of people do not like to buy lottery ticket, and they are probably not willing to buy flight insurance either. The overlapping of these two groups of people could be interesting for marketing purposes.

## 6. Implications of comparative expected utility theory

The paper proposes a normative and quantitative method to compare multiple mutually exclusive decision alternatives. The alternative decision theory aims to answer the questions that the other theories have failed to answer or with quantitative manners. The CEUT approach is designed based on classical expected utility theory. CEUT makes full use of the benefits of EUT, and develops its own process and definitions to implement evaluations. The theory is only designed to compare mutually exclusive decision alternatives. The application procedure and validations have been explained in previous sections. In the section, some important aspects of CEUT will be concluded and explained.

### 6.1. Subjective nature

CEUT is introduced with a normative approach and is expected to add rationality for decision making. However, the application of the approach needs subjective assignment of CCC elements. This process could vary from one decision maker to another. This is different from classical expected utility theory. However, the subjectivity during decision making is natural (Ravitch, 1989). Subjectivity should not be considered as a barrier from implementing rational decision approaches. Many theories have tried to suggest a universal template (either descriptive or normative) to analyze the different behaviors from different decision makers with different risk profiles. Under the scheme of comparative expected utility theory, different decision makers might get different answers by following the same procedures on the same decision problem. This may sound a bit random, but the answer shall be rational and consistent for each specific individual or group of people.

### 6.2. Quantitative approach

CEUT is introduced as a quantitative method. Probabilities and utilities are used to implement the comparisons. As is explained in Section 2.2, the assignment of CCC elements is subjective and might vary quite differently among different decision makers. For example, if CEUT results are to be used in budget planning or other purposes, the consistency of these financial numbers is impossible to maintain, as different people hold different opinions on the same outcome. Besides, the possibility of not including another possible decision alternative might have easily changed the CEUT results, because CEUT heavily depends on other decision alternatives. This will also make budget planning very dynamic and less convincing. The author will only recommend the CEUT approach as a quantitative method of comparing multiple mutually exclusive decision alternatives.

### 6.3. Axioms of CEUT

The evaluation and validation process of possible axioms of CEUT is shown in Section 2.5. The axioms of CEUT include:

- Dominance
- Transitivity of preference
- Invariance

The CEUT approach is a quantitative, and the dominance and transitivity axiom allow the decision makers to compare different decision options right by using the CEU values. These two axiom are important for a rational decision-making process. The dominance axiom also makes it straightforward to pick out the best alternative based on the theory. Invariance axiom brings consistency into decision-making process, and it should fulfill the need of a decision theory that could truly eliminate the influences from the different framings of the same decision question.

### 6.4. Relationship with EUT

There is a possibility that some decision makers would not care about the influence from other decision alternatives no matter what the possible outcome might be, and all their concern is the mathematical expected values for each options. This type of decision makers are not influenced psychologically by other decision options. Under this condition, EUT will become a special form of CEUT. No CCC elements will be identified. What axioms CEUT have will also be included in EUT, but EUT might have specific axioms that might not fit into CEUT. It is also possible that EUT might still keep all its own axioms defined by Von Neumann and Morgenstern (1944) and Savage (1972) based on this assumption.

## 7. *Future scope of work*

The potentials of comparative expected utility theory have been elaborated in the above. Some future work is hereby identified based on the discussions, including:

- More evidences are needed to further validate the alternative decision theory.
- Application of CEUT on decision alternatives with continuous probability distributions.
- Does the past or current statuses of decision makers (wealth, etc.) have influences on the decisions? And how does this change the way to use CEUT?
- If time aspects are included in the decision making, how could CEUT be adapted to make a fair comparison?
- How could CEUT be used for infinite expected value conditions, like St. Petersburg paradox (Bernoulli, 1954).